\documentclass[conference]{IEEEtran}
\usepackage{cite}
\usepackage{amsmath,amssymb,amsfonts}
\usepackage{algorithmic}
\usepackage{graphicx}
\usepackage{textcomp}
\usepackage{xcolor}
\def\BibTeX{{\rm B\kern-.05em{\sc i\kern-.025em b}\kern-.08em
    T\kern-.1667em\lower.7ex\hbox{E}\kern-.125emX}}
\begin{document}

\title{Modeling Gender Differences in Membership Change in Open Source Software Projects\\
{\footnotesize
}
}

\author{\IEEEauthorblockN{Olivia B. Newton}
\IEEEauthorblockA{\textit{College of Engineering and Computer Science} \\
\textit{University of Central Florida}\\
Orlando, FL, USA \\
olivianewton@knights.ucf.edu
}
\and
\IEEEauthorblockN{Jihye Song}
\IEEEauthorblockA{\textit{College of Engineering and Computer Science} \\
\textit{University of Central Florida}\\
Orlando, FL, USA \\
chsong@knights.ucf.edu
}

}

\maketitle

\begin{abstract}

Gender diversity in open source software development continues to be a topic of growing interest among researchers, practitioners, and organizations. 
To date, research has revealed disparities in participation between developers on the basis of gender, with women being significantly underrepresented in open source development. 
Using a large data set curated for studies of diversity in open source projects, we contribute to this body of work by characterizing the relationship between gender-based participation differences and group composition in GitHub.
In this paper, we present the results of a study examining the following: (1) similarities and differences between open source project contributors whose gender was inferred and those whose gender could not be inferred using existing gender inference tools, (2) the representativeness of a sample of open source projects with an identifiable mixed-gender contributor base to the population from which it was drawn (collaborative open source projects on GitHub), (3) differences between that sample of projects and a comparable sample of projects that did not have an inferable mixed-gender contributor base, and (4) a case study of participation dynamics in a small set of open source projects with a mixed-gender group of contributors. 
We found that contributors identified as women and contributors of unknown gender have a shorter tenure in open source projects compared to those identified as men. 
Additionally, at the team level, we found that project teams with mixed-gender composition were associated with lower turnover and teams with greater disparity in the distribution of platform tenure were associated with higher turnover. 
Finally, our case study reveals that when looking at the entire team, GitHub users were more likely to remain in a project rather than leave after contributing but when looking at women specifically, users were more likely to leave rather than remain after contributing to a project.
\end{abstract}


\section{Introduction}

Research finds that participation among women in open sourc software development is low compared to men \cite{el_asri_where_2019}.
This issue is more complicated than a lack of women who are interested in and skilled enough to participate in open source projects. 
Recent studies provide evidence that women face particular challenges and biases when attempting to participate \cite{balali_newcomers_2018, imtiaz_investigating_2019}. 
In the context of expertise and turnover in open source projects, the data suggest that this occurs even when they are more competent than their male peers \cite{terrell_gender_2017}, and that women are likely to disengage from open source projects faster than men \cite{qiu_going_2019}. 
Owing to these differences in participation, and because of the more general problem of diversity in STEM fields, group diversity has become an increasingly popular topic of discussion across technology-centered fields and also specifically in human-computer interaction (HCI) research \cite{himmelsbach_we_2019} and open source project research \cite{el_asri_where_2019, terrell_gender_2017, aue_social_2016, blincoe_perceptions_2019, daniel_effects_2013, ortu_how_2016, vasilescu_filkov_perceptions_2015}. 
Germonprez and colleagues suggest open source development is undergoing a transformation that may reflect ``an evolution or a coming crisis in how open source projects are able to encourage skillful, diverse, inclusive global work'' \cite[p. 4]{germonprez_rising_2019}.
We recognize this transformation as an opportunity to effect change, and aim to contribute to the related body of research through a study characterizing the factors that distinguish turnover in projects that are developed by mixed-gender groups and projects that are developed by single gender groups. 
To do this, we analyzed variable relationships extracted from a longitudinal data set curated by Vasilescu et al. \cite{vasilescu_serebrenik_data_2015} to promote studies of diversity in open source projects. 
Rather than focus solely on the lack of contributors who are not men in open source projects, we also pay particular attention to those projects that have a mixed-gender contributor base to characterize the participation dynamics observed within them. 
Additionally, we extend prior work on diversity and participation through an analysis of differences and similarities between open source project contributors whose gender is inferable via cues in a virtual space and contributors whose gender is not inferable due to the lack of such cues, or the noisiness of existing cues. 
Little research on membership change and group composition in open source projects in social coding platforms has explored this topic to date and it leads us to consider the limitations of research on gender in social coding platforms, both in terms of the lines of inquiry that are available using data extracted from social coding platforms and the questionable nature of studying related phenomena through binary categories of gender \cite{Keyes_misgender_2018, Steinhardt_fem_2015}.

\section{Turnover and Gender in Open Source Projects}

Membership change has been studied extensively by organizational and group researchers to understand the phenomenon and its effects on collaboration in a variety of work domains. 
The term membership change is used to describe the departure of members from a group, the addition or arrival of new members to a group, and, to a lesser extent, the retention of members in a group. 
Membership change is typically used interchangeably with the term turnover, although the latter term is often used to describe the rate at which group members leave and are replaced by newcomers \cite{Shaw1998}. 
Research on membership change in open source projects spans across a number of topics, from studies of onboarding and related social and technical challenges \cite{balali_newcomers_2018, steinmacher_overcoming_2019}, to the characterization of participation among different types of contributors \cite{barcomb_uncovering_2018}, to the prediction of turnover and its relation to outcomes like knowledge loss \cite{nassif_revisiting_2017}, software quality \cite{foucault_impact_2015}, and productivity \cite{mockus_organizational_2010}. 
In the following sections, we briefly summarize key findings from these studies and then describe our study of turnover and gender in GitHub. 

\subsection{Joining Projects}

Emerging in the early 2000s, research on onboarding in open source projects is extensive and, at least in part, driven by the observation that joining and specialization are both significant challenges for individuals interested in participation in software development and the individuals and teams that mentor them. 
Some early work emphasizes the types of behavior---referred to as "joining scripts"---that newcomers should engage in when attempting to join a project community \cite{von_krogh_community_2003}. 
These joining scripts are based largely on activity; newcomers are expected to maintain some level of activity and engage in particular types of activity (e.g., reporting bugs) if they would like to be accepted into a project’s community.

Recent work takes a nuanced approach to onboarding in open source projects, distinguishing between different types of barriers faced by newcomers and identifying ways that project managers may be able to lower barriers and support participation. 
Research in this area identifies technical and social challenges for newcomers attempting to join open source projects in addition to the challenges faced by project maintainers to identify and mentor candidates in the newcomer pool for particular tasks \cite{balali_newcomers_2018, gousios_work_2016, steinmacher_overcoming_2019}. 
In a study conducted by Balali and colleagues, some mentors expressed experiencing "difficulty in creating an inclusive community" \cite[p. 693]{balali_newcomers_2018}, specifically pointing to the issue of correct gender pronoun usage. 
Furthermore, contributors who are women shared that they felt "less comfortable with and accepted by their counterparts who are men" \cite[p. 702]{balali_newcomers_2018}. 
Taken together, these findings reveal a tension experienced by not only cis women\footnote{The descriptor cis, or cisgender, indicates an individual whose gender aligns with sex assigned at birth. A cis woman is thus one who was assigned female at birth and is a woman \cite{schilt_doing_2009}.}, but also contributors of other marginalized genders (e.g., trans women and men). 
This suggests there remains a set of challenges in defining and identifying contributors on the basis of gender and a gap in understanding with respect to the experiences and inclusion of these contributors in open source projects.

Pointing to the need to consider more tacit factors, research conducted by Fronchetti et al. \cite{fronchetti_what_2019} and Qiu et al. \cite{qiu_going_2019} models and characterizes the signals that social coding platform users rely on when choosing a project to contribute to from the options available to them.
Fronchetti et al. \cite{fronchetti_what_2019} constructed a model to investigate the factors that predict developer onboarding in open source projects and found that popularity (measured in stars), the time to merge pull requests, and number of programming languages were the highest ranked predictors; among these factors, popularity was the strongest predictor of developer onboarding. 
The effect of the number of programming languages in their modeling is consistent with prior theorizing on contribution barriers by von Krogh et al. \cite{von_krogh_community_2003}, in which developer familiarity with programming language contributes to task difficulty and can thus impede participation. 
Qiu and colleagues \cite{qiu_going_2019} employed a mixed-methods approach to identify the signals used by contributors in choosing projects, and characterized those signals by investigating their observability in social coding platforms. 
They found that the open source contributors rely on many signals when selecting projects to join: the level of activity in the project, the popularity of the project, the disposition of issue and pull request handlers, the presence of issue and pull request templates and labels, and the presence of an organized, detailed README. 
Last, in interviews with volunteer contributors, Barcomb et al. \cite{barcomb_uncovering_2018} found that contributions in open source projects are oftentimes the result of an invitation extended by a known person, providing evidence that social ties and norms play an important role in participation. 
In sum, the platform features influencing participation range from implicit to explicit, where, for example, signals vary by the degree to which they are observable, or discoverable, with some signals necessitating multiple observations and/or actions on the part of the potential contributors. 
Moving beyond the platform, participation is sometimes driven by overt requests to join and may depend on the breadth of connections one has in a community. 
These studies thus uncover some of the factors influencing participation decisions, factors which, in some cases, may differ depending on gender.

\subsection{Leaving Projects}

Early research on turnover in open source projects suggested that, in these types of projects, membership change is quite common in the core group of contributors (i.e., the most active 20\% of all committers to the project \cite{robles_contributor_2006}). 
However, this finding is not consistent with network-based measures of developer role showing that the core group is highly stable whereas the periphery group is volatile \cite{joblin_apel_mauerer_evolutionary_2017, joblin_apel_hunsen_classifying_2017}. 
In studies of open source projects, high turnover rates are associated with lower software quality and productivity, and increases in defects as newcomers lack sufficient expertise and exhibit different levels of activity when compared to established contributors in the project \cite{foucault_impact_2015, mockus_organizational_2010, rigby_quantifying_2016}. 
Researchers have also found that both gender diversity (variety in the gender of a project’s contributors) and tenure diversity (variety in the tenure of a project’s contributors) in open source predict turnover in medium-sized and large teams \cite{vasilescu_posnett_gender_2015, vasilescu_serebrenik_data_2015}.

\subsubsection{Gender Differences}

HCI and software engineering researchers have examined gender differences in participation and the diversity of collaborations in open source projects. 
Qiu et al.'s \cite{qiu_going_2019} research demonstrates that there are quantifiable gender differences in the length of time before a contributor departs from a project, specifically showing that women are more likely to disengage before men. 
Some research has begun to explore the presence and effects of bias experienced in open source projects. 
This work is insightful given that open source developers have some level of awareness of potential collaborators’ gender in social coding platforms \cite{vasilescu_posnett_gender_2015} and the salience of gender can influence interactions in social spaces \cite{McNicol_noneya_2013}. 
Indeed, Terrell et al. \cite{terrell_gender_2017} found that contributions made by women whose gender is identifiable are accepted at lower rates when they are not "insiders" and that they are accepted at higher rates than those of men when their gender is not identifiable, indicating that negative evaluations of women's contributions are driven by bias, rather than quality of work. 
Imtiaz and colleagues \cite{imtiaz_investigating_2019} investigated the effects of different types of bias that may be experienced by women in addition to contribution differences between genders in social coding platforms. 
They found that the contributions of women were centralized to a smaller set of projects compared to men.
Furthermore, their results showed that, in communication, women were less likely to use profanity and less likely to express positive or negative sentiment, thus remaining relatively reserved in their interactions compared to men. 
Findings across the related areas of research suggest that the low participation of women in open source projects is a complex issue. 

\subsection{Present Work}
We contribute to this body of work in two ways. First, we conducted a big data analysis of GitHub projects to delineate the association between gender, contributor tenure and turnover, and project characteristics. Second, we conducted a case study of membership change in projects with a mixed-gender contributor based on fifteen projects.

Our big data analysis was focused on deriving general insights about gender differences in GitHub: (1) similarities and differences between contributors whose gender can be inferred and contributors whose gender remains obscured; (2) the representativeness of a sample of open source projects with a mixed-gender contributor base to the population from which it is drawn (i.e., collaborative open source projects on GitHub); (3) differences between that sample of projects and a comparable sample of projects that do not have an identifiable mixed-gender contributor base. The following questions guided the big data component of our research:

\begin{itemize}
    \item \textit{RQ1. Gender Differences in Platform Tenure}: Are there disparities in sustained participation between gender groups on GitHub?
    \begin{itemize} 
    \item \textit{RQ1.1}: Do women typically have shorter platform tenure compared to men? Based on research examining differences in tenure in software engineering firms \cite{james_what_2017} and open source projects \cite{qiu_going_2019}, we expect that women, on average, have shorter platform tenure when compared to men. 
    \item  \textit{RQ1.2}: Is the distribution of platform tenure among project contributors of an unknown gender similar to that of either women or men, or is it quantitatively distinct?
    \end{itemize}
    \item \textit{RQ2. Representativeness of Mixed-Gender Teams}: Are projects maintained by a mixed-gender contributor base quantitatively similar to or distinct from the broader open source project population?
    \item \textit{RQ3. Turnover in Mixed-Gender Teams}: Does the mixed-gender status of a project team predict differences in turnover? 
\end{itemize}

Our case study was focused on characterizing participation dynamics in projects maintained by a mixed-gender group of contributors by modeling the probability that contributors, overall and grouped by gender, will join a project, remain present in a project, and/or maintain a period of absence in the project after initial participation. The following questions guided the case study component of our research:

\begin{itemize}
    \item \textit{RQ4. Representativeness of Case Study Sample}: Is the case study sample quantitatively similar to or distinct from the broader open source project population on GitHub? 
    \item \textit{RQ5. Project Contributor Transitions}: Are contributors more likely to remain in a project after joining? We expect that, overall, contributors have a greater probability of staying with a project than leaving a project after joining \cite{joblin_apel_mauerer_evolutionary_2017}.
    \item \textit{RQ6. Gender Differences in Contributor Transitions}: Do transition probabilities differ for contributors on the basis of gender? We expect that men have a higher probability of staying with a project rather than leaving after joining \cite{qiu_going_2019}. We expect that women have a higher probability of leaving a project rather than staying after joining \cite{qiu_going_2019}.
\end{itemize}

\section{Method}

For our research study, we used the data set curated and shared by Vasilescu and colleagues \cite{vasilescu_serebrenik_data_2015}. 
To explore differences in sustained participation on the GitHub platform, we first analyzed platform tenure across three groups, with contributors labeled as: woman, man, or unknown. 
We then analyzed a subset of the data to identify predictors of membership change on the GitHub platform for two samples drawn from the population via mixed effects models. 
From the subset of projects with a mixed-gender contributor base, we selected a small number of projects to model participation dynamics. 
In this section, we first describe Vasilescu et al.’s \cite{vasilescu_serebrenik_data_2015} data set, including the computation and definition of their variables. 
We then detail the approach used to sample projects and construct models for the prediction of membership change (big data) and the characterization of participation dynamics (case study).

\subsection{Data Set}

The longitudinal data set of 23,493 GitHub projects and 122,014 users was collected and enhanced by Vasilescu et al. \cite{vasilescu_serebrenik_data_2015} with the goal of enabling studies of diversity in open source projects on GitHub, a platform that provides tools and services to support software development and its management \cite{begel_social_2013}. 
Projects were selected from the GHTorrent data dump 1/2/2014 for inclusion by the researchers if the project had at least 2 committers, 10 total commits, and 6 months of history. 
These selection criteria allowed for the curation of a data set corresponding to active, collaborative open source projects hosted on GitHub, where a project is defined as a base repository and all of its forks, or copies of the base. 
The data set includes information about both GitHub users (e.g., gender, commit activity, etc.) and projects (e.g., age, main programming language, number of watchers, etc.).

To enhance the data set, Vasilescu et al. \cite{vasilescu_serebrenik_data_2015} applied a username aliasing approach to user data in order to identify project contributors using multiple aliases and merged their information. 
In addition to this, the researchers used a gender resolution technique to infer the gender of project contributors. 
This technique leveraged name and location data to probabilistically determine gender and was used to infer the gender of 873,392 users ("32.6\% of all users, but 80\% of those who disclosed their names" \cite[p. 515]{vasilescu_serebrenik_data_2015}). 
Vasilescu et al. reported that, for these users, 91\% were labeled as men and 9\% were labeled as women; the gender of the remaining users was labeled as unknown. 
Specifically, gender was inferred using the genderComputer tool \cite{vasilescu_human_2014} which uses the individual’s name and location in combination with “transformations, diminutive resolution, and heuristics” to infer gender \cite[p. 515]{vasilescu_serebrenik_data_2015}. 
User and project data were segmented into quarters (i.e., 3-month periods). 
1,136 projects in the data set did not have a main programming language and were excluded from further processing, resulting in the selection of 22,357 open source projects.

\subsection{Big Data: Predictive Modeling}

We modeled turnover in projects with a mixed-gender contributor base (i.e., those with at least one woman on the team) and projects that did not have an mixed-gender contributor base (i.e., those where there were no identified women on the team). 
To do this, we selected a subset of the projects in the data set if, and only if, there was at least one contributor who was a woman in at least one quarter of the project. 
From the 23,357 projects in the data set, only 5,539 had at least one woman contributor and were labeled as mixed-gender teams; the remaining 16,818 were labeled as not mixed-gender team. 
These two groups are unbalanced and, in the interest of comparing teams with similar quantitative characteristics to better understand the relationship between gender and membership change, we applied a sampling technique to select 5,539 projects from the subset of 16,818 projects that did not have women on the team. 
We next describe this sampling approach.

\subsubsection{Sample Selection}

To attain equal samples for the mixed-gender team categorical variable, we constructed a vector for each project and calculated the euclidean distance between them to select quantitatively similar projects. 
We created a vector of features for each project that consisted of: number of contributors (12-month period), number of commits (12-month period), number of forks, number of watchers, and project age. 
First, we normalized the feature vectors so that each feature's value was in a range between zero and one. 
The normalization of these features prevented bias toward features with a larger range. 
Then, for every project labeled as a mixed-gender team, we found the nearest project—where the distance between two projects is the euclidean distance between their feature vectors—that did not have a mixed-gender team. 
This allowed us to reduce the set of 16,818 projects that did not have a mixed-gender team to a set of 5,539 projects. 
The final sample for membership change prediction modeling contained 11,078 projects with an even split across the mixed-gender team variable.

\subsubsection{Sample Evaluation} 

Following the guidance of Nagappan et al. \cite{nagappan_diversity_2013}, we evaluated the representativeness of the samples used in our analyses.  
To do this, we employed their vocabulary and technique for measuring sample coverage, specifically using their algorithm as implemented in statistical computation software R \cite{r_core_team_2019}. 
This vocabulary and technique was proposed with the aim of improving the generalizability of methods and findings in software engineering research. 
As a first step, the universe, or population, is defined and projects in that universe are then characterized along one or more dimensions. 
The set of dimensions is selected on the basis of their relevance to the research topic; these dimensions "define the space of the research topic" within the universe \cite[p. 2]{nagappan_diversity_2013}.

For our purposes, the population consists of the open source projects developed by a team (i.e., more than one person) on GitHub at the time of data collection; that is, all of the projects in Vasilescu et al.'s data set. 
For empirical research on membership change, we contend that, at minimum, the space consists of the following dimensions: number of contributors (12-month period), number of commits (12-month period), number of forks, number of watchers, main programming language, and project age. 
The selection of these dimensions is based on prior research showing that activity and popularity are related to growth and attraction of newcomers \cite{fronchetti_what_2019}.

\subsubsection{Model Implementation} 

We constructed mixed effects models to analyze the factors that predict membership change in open source projects. 
These models were implemented using statistical computation software R \cite{r_core_team_2019} and, in particular, the lmer function in the lme4 package \cite{bates_lme4_2015}. 
The following variables were implemented as fixed effects:

\begin{itemize}
    \item \textit{Mixed-gender team}: binary value representing the presence of women in the contributor group across project lifetime.
    \item \textit{Team size}: number of contributors (committers, pull request submitters, commenters) in a given quarter.
    \item \textit{Tenure disparity}: calculated using the Gini coefficient, a measure of disparity [0,1], where higher values indicate greater disparity in contributors' tenure.
    \item \textit{Pull requests}: number of pull requests in a given quarter.
    \item \textit{Comments}: number of comments in a given quarter.
    \item \textit{Issues}: number of issues in a given quarter.

\end{itemize}

Ratio of turnover served as the response variable and we included the intercept of a project as a random effect. 
The \textit{p}-values reported were obtained using the likelihood ratio test with the anova function from the stats package. 
For the models, which included a single fixed effect, the test compared the full model with the particular fixed effect against a null model without the particular fixed effect. 
All assumptions of the models were checked by examination of residual plots.

The primary fixed effect of interest was group composition with respect to gender; that is, was the project team labeled as mixed gender or not.
In addition to team mixed-gender status, we modeled the effects of team size, platform tenure disparity, and three indicators of project activity (pull requests, comments, and issues). 
For team size, the project team consisted of GitHub users who contributed to the project, including committing code, submitting pull requests, contributing to discussion via comments, and reporting issues \cite{vasilescu_serebrenik_data_2015}. 
Several variables representing different types of tenure were provided in the data set. 
We only made use of the GitHub tenure variable as it was most appropriate for our research question concerning participation in the broader open source development taking place in GitHub rather than in a specific project. 
Platform tenure disparity was included in our modeling because we contend that it reflects important differences in group composition with respect to status and experience based on activity traces \cite{marlow_impression_2013}, and has implications for knowledge management and turnover \cite{newton_expertise_2019}. 
Tenure disparity was calculated for each project team at each quarter using the Gini coefficient. 
The Gini coefficient is a calculation initially developed to study income disparity \cite{Dorfman_1979}, and more recently adapted to study disparities in other domains (e.g., in group diversity research \cite{Solanas_2012}). 
The larger a Gini coefficient is, the higher the centralization among a small number of people in a population. Groups with a high value have higher levels of disparity in project tenure (e.g., many relative newcomers and few contributors with long tenure).

\subsection{Case Study: Transition Probabilities}

In this section, we describe the approach we used to model the dynamics of membership change in a set of projects selected from the mixed-gender team sample.
Nagappan et al. \cite{nagappan_diversity_2013} developed a greedy algorithm for the selection of projects that maximizes the coverage of a sample. 
We used their algorithm to select fifteen projects from the mixed-gender projects sample and use their sample coverage scoring method to evaluate its representativeness. 

\subsubsection{Contributor Transitions}

We adopted an approach used in prior work by Joblin and colleagues \cite{joblin_apel_mauerer_evolutionary_2017} to examine the likelihood of open source contributors joining and leaving a project.
Joblin et al. used sequential data modeling to evaluate the stability of the contributor base in ten large open source projects. 
Specifically, they used the discrete state Markov model in which they assigned a state to developers at each time window for a project. 
We similarly assigned one of two discrete states to project contributors for each quarter of the project lifetime. 
In each quarter, a project contributor is classified as either absent or present; a contributor was assigned an absent value for each quarter prior to the quarter that they first joined the project and for each quarter after they left the project. 
A contributor was assigned a present value for each quarter that they engaged in activity in the project as recorded in the GitHub data.
For this classification, we used four variables computed by Vasilescu and colleagues: team, which lists the user IDs of contributors who were considered members of the project team; left, the list of users who left the project in a given quarter; joined, the list of users who joined a project in a given quarter; and stayed, the list of users who remained in the project in a given quarter. 
We thus represented project contributor state transitions as a transition matrix where each row corresponded to a single contributor and each column corresponded to a specific quarter. 

\section{Results}

\begin{table}
\renewcommand{\arraystretch}{1.1}
  \caption{User count and GitHub tenure by gender.}
  \label{tab:userCount}
  \begin{center}
  \begin{tabular}{ l c c }
  \hline
    \textbf{Gender} &
    \textbf{User Count} (\%) & 
    \textbf{Median GitHub Tenure} \\
    \cline{1-3} 
    \textit{Women} & 5,284 (7\%) & 502 \\
    
    \textit{Men} & 57,103 (74\%) & 549 \\
    
    \textit{Unknown} &  14,813 (19\%) & 506 \\
    \hline
\end{tabular}
\end{center}
\end{table}

\subsection{Big Data: Gender Differences in Platform Tenure}

\begin{figure}
  \centering
  \includegraphics[width=1\linewidth]{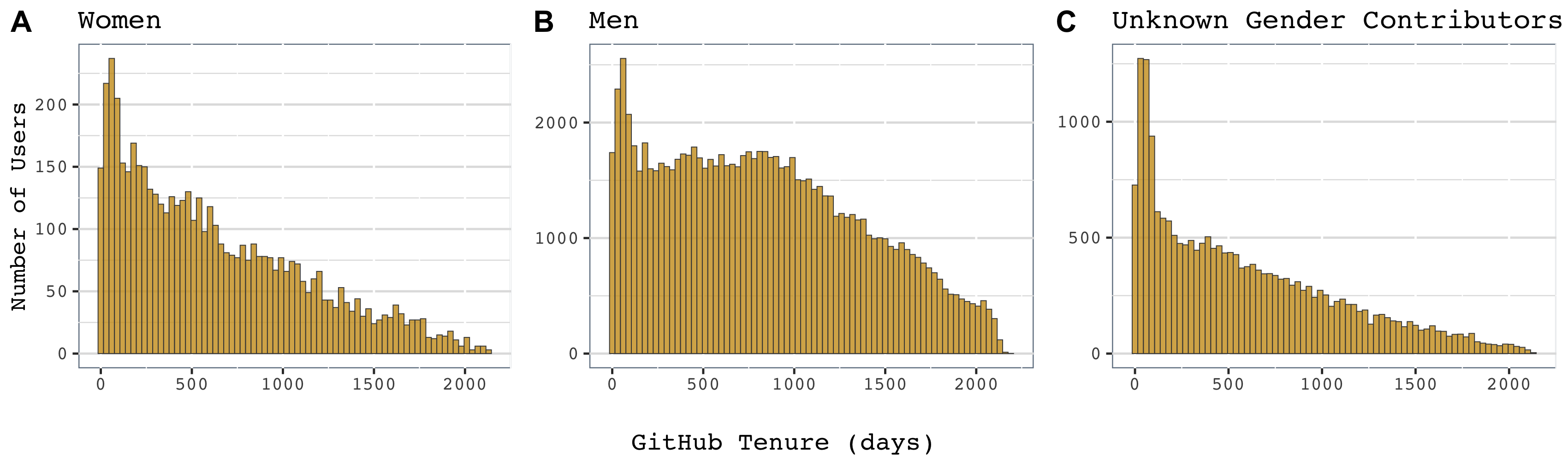}
  \caption{The distribution of platform tenure for GitHub users in Vasilescu et al.'s data set. Most women and unknown gender contributors have relatively short tenure on the platform.
  }
  \label{fig:tenure}
\end{figure}

Consistent with findings in research on software engineering organizations \cite{james_what_2017}, in Vasilescu et al.'s data set, we observed that women tend to have shorter tenure in open source projects when compared to men (Table \ref{tab:userCount}).
This observation complements work by Qiu et al. \cite{qiu_going_2019} showing that women were more likely to disengage from the platform sooner than men. 
Furthermore, we observed that contributors of unknown gender similarly have shorter tenure when compared to contributors who were identified as men. 
Distribution plots show that, compared to both contributors identified as women and of unknown gender, men are more heavily represented across levels of platform tenure (Figure \ref{fig:tenure}).
These plots also show that the distributions of platform tenure for women and contributors of unknown gender are similar; both distributions have right skew. 

To statistically assess differences between these groups, Kruskal-Wallis tests, non-parametric equivalent of a one-way analysis of variance, were applied to users' platform tenure data (Table \ref{tab:Kruskal}). 
Women and unknown gender contributors had a lower median tenure compared to men in the platform, and there was an overall significant difference between groups p $<$ .0000. 
There was a significant difference between women and men in addition to a significant difference between men and contributors of unknown gender.
The difference between women and contributors of unknown gender was however non-significant. 
These results suggest that few women joined the GitHub platform at its inception and/or some of the women who joined early on did not remain on the platform for very long. 
Contributors of unknown gender in the sample exhibit a similar pattern of engagement which may be evidence of other similarities between these groups (i.e., unknown gender group is made up people who face a particular set challenges and biases in open source projects).  

\begin{table}
\renewcommand{\arraystretch}{1.1}
  \caption{Results of Kruskal-Wallis tests. GitHub platform tenure is counted in days; the difference in medians is thus counted in number of days for the two groups being compared.}
  \label{tab:Kruskal}
  \begin{center}
  \begin{tabular}{ l c c c }
    \hline
    \textbf{Groups} &
    \textbf{Difference} &
    \textbf{\textit{H}-value} &
    \textbf{\textit{p}-value} \\
    \cline{1-4}
    \textit{Women and Men} & 47 & 1224.80 & .0000 \\
    \textit{Women and Unknown Gender Users} & 4 & 3.11 & .08 \\
    \textit{Men and Unknown Gender Users} &  43 & 3198.10 & .0000 \\
    \hline
\end{tabular}
\end{center}
\end{table}

\subsection{Big Data: Modeling Membership Change}

The summary statistics for the two groups in the sample---all projects with identifiable mixed-gender contributor base and the sample of projects that did not have an identifiable mixed-gender contributor based---are provided in Table \ref{tab:sampleSum}. 
To aid in the evaluation of our sample coverage, we first characterize the population along the relevant dimensions. 
The primary language of most projects was JavaScript.
Additionally, most projects had a relatively small contributor base and few forks and watchers but were at least a year old. 
For the samples used in our mixed effects models, we applied the sample coverage algorithm to score the samples drawn from the population. 
The coverage scores for these samples are provided in Table \ref{tab:coverage}.

\begin{table}
\renewcommand{\arraystretch}{1.1}
  \caption{Mean values for two groups in sample for mixed effects model. Median values are identical for both groups as a result of sampling technique described in Methods: Sample Selection.}
  \label{tab:sampleSum}
  \begin{tabular}{ l c c}
    \hline
    ~ &
    \textbf{has\_woman = TRUE} & \textbf{has\_woman = FALSE}
    \\
     ~ & All Projects & Sample
     \\
     \cline{1-3}
    \textit{Contributors (12mos)} & 19.23 & 9.05 \\
    \textit{Commits (12mos)} & 18,762.70 & 1,064.64 \\
    \textit{Forks} & 25.76 & 10.58 \\
    \textit{Watchers} &  52.57 & 24.70 \\
    \textit{Project Age} & 16.93 &  17.11 \\
    \hline
\end{tabular}
\end{table}

\begin{table}
\renewcommand{\arraystretch}{1.1}
  \caption{Coverage scores for samples from study population. }
  \label{tab:coverage}
  \begin{center}
  \begin{tabular}{ l c c }
    \hline
     &
    \textbf{Mixed-Gender} & 
    \textbf{Not Mixed-Gender} \\
    ~ & All Projects & Sampled Projects \\
    \cline{1-3}
    \textbf{\textit{Score}} & 0.881 & 0.885 \\
    \textit{Main Language} & 0.998 & 0.997 \\
    \textit{Contributors (12mos)} &  1 & 0.998 \\
    \textit{Commits (12mos)} &  1 & 1 \\
    \textit{Forks} &  1 & 1  \\
    \textit{Watchers} &  1 & 1 \\
    \textit{Project Age} &  1 & 1 \\
  \hline
\end{tabular}
\end{center}
\end{table}

\subsubsection{Model Results}

Table \ref{tab:singTermModels} provides a summary of the models, including model estimates and standard error (se), t-values, and R$^2$ values. 
Although each model was statistically significant, the models with mixed-gender team status and platform tenure disparity as terms have the largest estimates, and the former has the largest R$^2$ value. 
These results suggest that that group composition, both in terms of gender and the distribution of status and expertise in a project are associated with differences in turnover in open source projects. 

\begin{table}
\renewcommand{\arraystretch}{1.1}
  \caption{Summary of results for mixed effects models. Model includes term as fixed effect and project as random effect.}
  \label{tab:singTermModels}
  \begin{tabular}{l c c c c}
    \hline
    \textbf{Model Term} & 
    \textbf{Estimate} $(\pm se)$ & \textbf{\textit{t}-value} & 
    \textbf{\textit{$R^2$}} &
    \textbf{\textit{$\Delta R^2$}} \\
    ~ & ~ & ~ & ~ & \tiny from null mod \\
    \cline{1-5}
    mixed\_gender\_team & -0.020 ($\pm$ 0.003) & -5.17**** & .27 & -.00  \\
     Gini\_gh\_tenure & 0.160 ($\pm$ 0.007) & 23.40**** & .25 & -.02  \\
    num\_team & -0.001 ($\pm$0.000) & -19.59**** & .25 & -.02 \\
     num\_pull\_req & -0.000 ($\pm$ 0.000) & -10.61**** & .26 & -.01  \\
    num\_comments & -0.000 ($\pm$ 0.000) & -20.32**** & .25 &  -.02 \\
    num\_issues & -0.002 ($\pm$ 0.000) & -24.35**** & .23 &  -.04 \\
  \hline
  \tiny Note: 
  **** \textit{p} $<$ .0001
\end{tabular}
\end{table}

\subsection{Case Study: Modeling Contributor Transitions}

The sample coverage score results for the fifteen projects that were selected are provided in Table \ref{tab:CScoverage}. 
Overall, the sample has low coverage. 
This is expected given the small size of the sample relative to the large population size. 
Additionally, the low score appears to be driven primarily by the main language and project age dimensions.

\begin{table}
\renewcommand{\arraystretch}{1.1}
  \caption{Coverage scores for case study sample.}
  \label{tab:CScoverage}
  \begin{center}
  \begin{tabular}{ l c }
    \hline
     &
    \textbf{Case Study Sample}  \\
    \cline{1-2}
    \textbf{\textit{Score}} & 0.044  \\
    \textit{Main Language} & 0.691  \\
    \textit{Contributors (12mos)} & 0.997  \\
    \textit{Commits (12mos)} & 0.956 \\
    \textit{Forks} & 0.997 \\
    \textit{Watchers} & 0.991 \\
    \textit{Project Age} & 0.644 \\
  \hline
\end{tabular}
\end{center}
\end{table}

\begin{table}
\renewcommand{\arraystretch}{1.1}
  \caption{Team size and programming language information for projects in case study sample.}
  \label{tab:CSsummary}
  \begin{center}
\begin{tabular}{ c c c }
    \hline
    \textbf{Project} & \textbf{Team Size} & \textbf{Main} \\
        ~ & (\# Women) & \textbf{Language} \\
    \cline{1-3}
    \textbf{1} & 145 \small (6) & Ruby \\
    \textbf{2} & 38 \small (2) & Python \\
    \textbf{3} & 45 \small (2) & JavaScript \\
    \textbf{4} & 47 \small (5) & Java \\
    \textbf{5} & 16 \small (1) & Python \\
    \textbf{6} & 35 \small (3) & JavaScript \\
    \textbf{8} & 112 \small (9) & C \\
    \textbf{9} & 35 \small (2) & Python \\
    \textbf{10} & 25 \small (2) & Python \\
    \textbf{11} & 15 \small (2) & Ruby \\
    \textbf{12} & 47 \small (1) & Ruby\\
    \textbf{13} & 39 \small (1) & JavaScript\\
    \textbf{14} & 98 \small (5) & C++ \\
    \textbf{15} & 129 \small (3) & C++ \\
  \hline
\end{tabular}
\end{center}
\end{table}

\subsubsection{Contributor Transitions}

\begin{figure}
\centerline{\includegraphics[width=1\linewidth]{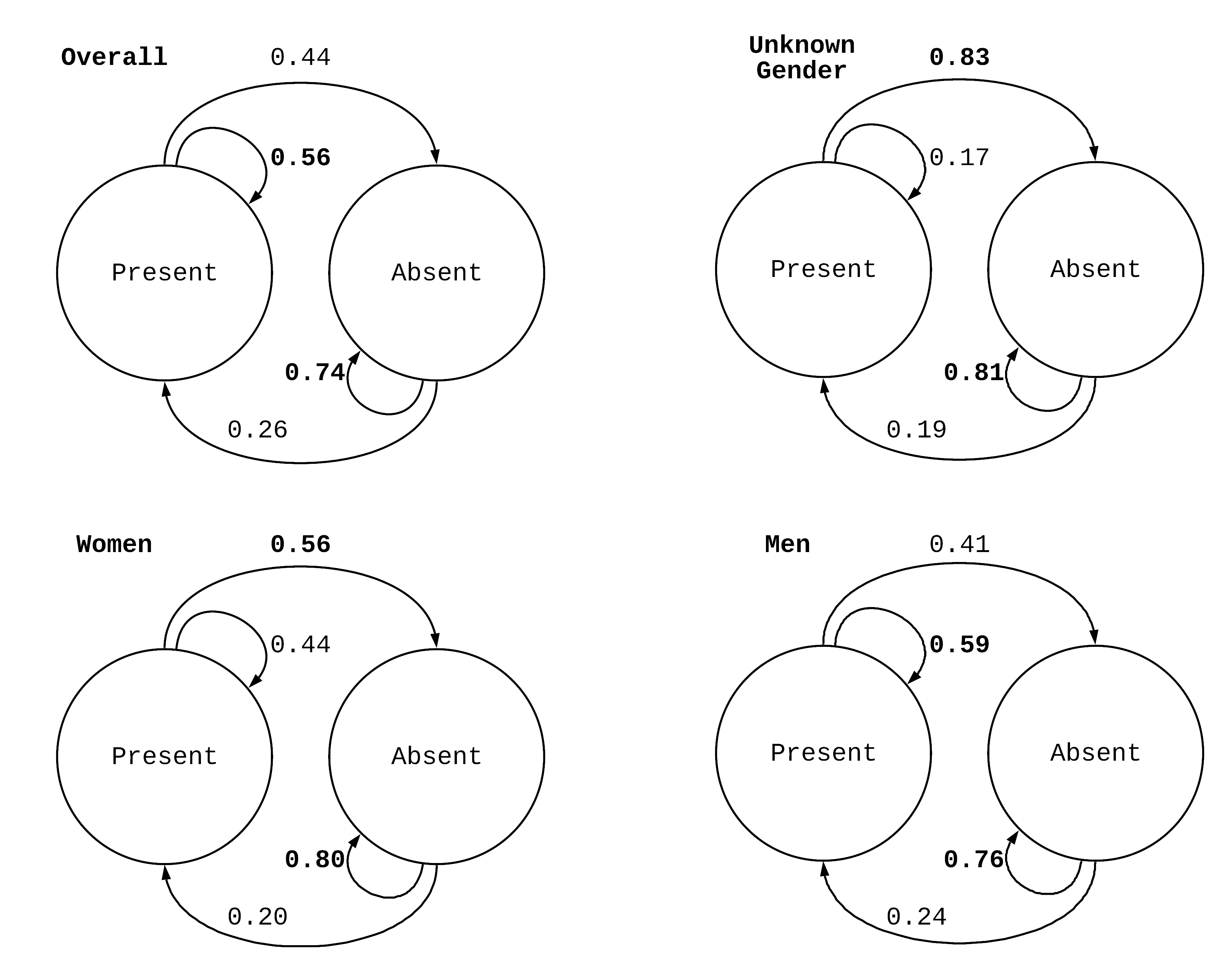}}
\caption{Participation dynamics in GitHub: transition probabilities averaged across all project contributors in case study sample and by gender group.}
\label{fig:transprob}
\end{figure}

Median values for transition probabilities for users in the case study sample are given in Figure \ref{fig:transprob} and details about projects in Table \ref{tab:CSsummary}.
Overall, we observed some stability in these projects: contributors had a higher probability of remaining in the project after joining (9 out of 15 projects), although there are some exceptions. 
When looking at the transition probabilities for the woman or group of women in a project, it was more likely that a contributor would leave a project after joining (i.e., probability of present $\rightarrow$ absent is greater than probability of present $\rightarrow$ present). 
This is in contrast to the transition probabilities for men, who were more likely to remain in a project after joining (i.e., probability of present $\rightarrow$ present is greater than probability of present $\rightarrow$ absent). 
This suggests the group of men in mixed-gender teams is more stable than the group of women in mixed-gender teams. 
However, relevant to the difference between the overall trend and the trend among women, all of the projects included in the case study had teams that were made up primarily of men (see Team Size column in Table \ref{tab:CSsummary}). 
Even in the two projects with smaller teams, projects 5 and 11, there were few women, 1 and 2 respectively. 
It is possible that the observation that women had a greater probability of leaving rather than remaining in these projects is due, in part, to the reported harassment and lack of comfort experienced by women contributors when working alongside men in open source projects \cite{balali_newcomers_2018, terrell_gender_2017, nafus_patches_2012}. 
These results in addition to the finding that women’s contributions tend to be centralized to a smaller number of projects than those of men \cite{imtiaz_investigating_2019} help explain the shorter tenure of women in open source projects and the GitHub platform in general. 

\section{Discussion}

In this study we set out to study factors that are related to the low participation of women in open source projects. 
We found that, in addition to factors already linked to turnover in open source projects, group composition in terms of gender predicted levels of turnover with mixed-gender teams exhibiting lower levels. 
The significant effect of gender composition on turnover complements findings from a study by Blincoe et al. \cite{blincoe_perceptions_2019} in which developers reported more negative behaviors and experiences in male-only teams.
In other words, the unpleasant nature of such experiences in homogeneous teams produces increases in turnover. 
We analyzed projects and their composition to better understand similarities and differences between contributors based upon gender.
Our findings add to the body of evidence showing that participation dynamics vary by gender.
This research also extends prior work by revealing similarities in tenure between GitHub users who are women and those whose gender is unknown.
Through our case study, we demonstrated the utility of transition-probability models for examining group differences in contexts where participation is ephemeral yet recurrent. 
The application of this approach in our study showed that women are not only more likely to leave a project compared to men, but that they are also less likely to return to a project after a period of absence. 
Furthermore, the participation of GitHub users of unknown gender in a project is very limited---they contribute in a small period of time and their departure from the project is more permanent.

In conducting this study, we considered the ways that the design of the GitHub platform shapes studies of gender in open source projects. 
First, the platform does not request gender information upon account creation and, as a result, likely diminishes the salience of gender and alters its effects on social interactions \cite{McNicol_noneya_2013}. 
This reflects a particular design choice and, embedded within it, assumptions about the types of and ways that individuals use the platform that have significant implications, both for the phenomena of interest and the ways that the phenomena can be studied. 
Although we do not make a claim to the appropriateness of this design choice, we do contend that it is one that warrants further attention. 
Second, this design choice results in the need for researchers interested in analyzing differences and similarities between genders on the platform to infer gender based on a set of, arguably, noisy cues, including, for example, displayed name, location, and, as some other researchers have used, picture and email information to locate accounts on other platforms that belong to the same user which can result in misgendering \cite{Keyes_misgender_2018}.
Related to this, and how gender information is requested or otherwise inferred, we reflected on the observation that research on gender in open source projects has largely relied on binary categories of gender to account for differences and similarities in participation.
Platform design, in combination with the gender binary frame applied to study gender differences, has resulted in an analysis of genders as relatively homogeneous groups and the potential exclusion of open source contributors of other marginalized genders, including in the research presented here. 
This presents an opportunity for future research to better understand and characterize within group differences and different types of diversity in open source projects \cite{himmelsbach_we_2019}. 
Researchers can employ a mixed-methods approach to provide a more comprehensive analysis of the gender binary issue and gender categories.

Like other research on membership change in open source development, our study focused on modeling factors internal to projects (i.e., group composition and project activity). 
Less work has examined how factors exogenous to a specific project and relevant to the larger population or ecosystem may alter turnover within them. 
Significant changes to the platform and decisions made by corporate stakeholders may have the potential to instigate an exodus of users from the platform. 
For example, some GitHub users expressed dissatisfaction with the decision to sell the platform to Microsoft and claimed they would leave the platform following the acquisition \cite{warren_GHacq_2018}. 
GitHub has also been criticized by both its users and employees for their continued relationship with the U.S. Department of Homeland Security’s Immigration and Customs Enforcement (ICE) \cite{chan_microsoft-owned_2019, ghaffary_github_2019}. 
These actions taken by GitHub decision makers likely influence the choices of some developers to participate in open source development on the platform. 
Here, an understanding of within group differences could illuminate the causes of diversity issues in open source along other sociodemographic dimensions. 
Such an understanding however necessitates an analysis of how race, class and status intersect with gender in studies of membership change. 
Because these types of changes and decisions are published, future research can leverage information about the occurrence of these events to investigate their effects on membership change in open source projects. 

In the work presented here, we suggest GitHub users whose gender was not identifiable by Vasilescu were potentially not men, but also not necessarily women. 
While there were quantitative similarities between users who were identified as women and unknown gender users, additional research is needed to accurately characterize the latter group.
The group of users whose gender was not resolved may be made up individuals who generally exhibit lower levels of engagement in the platform for different reasons and thus do not provide the type of information that is used to infer gender by researcher.
The reasons for lower levels of engagement may be tied to gender (e.g., the bias and harassment \cite{nafus_patches_2012}), or other factors that are not necessarily specific to gender but reflect other sociodemographic dimensions (e.g., limited time or financial resources for ongoing participation).

\subsection{Limitations}

The data set used for the analysis presented here was collected by Vasilescu et al. from a GHTorrent data dump in 2014. 
It does not capture changes in the platform user base and participation in open source projects that have occurred since then. 
This limits the interpretation of results. 
For example, it is possible that there has been a reduction in the gender imbalance observed in GitHub. 
Recent studies of gender differences and diversity suggest, however, that this is likely not the case (e.g., \cite{el_asri_where_2019, qiu_going_2019}). 
We therefore expect the application of our analyses to a more recent data set would likely produce similar findings.
The interpretation of results is also limited given the inability to accurately label gender for all users.
It is possible that a project classified as having a mixed gender contributor base could have a team size of two, and that one contributor be labeled as a woman and the other unknown.
Furthermore, it is possible that the user of unknown gender is a woman, meaning the project is not maintained by a mixed gender team.
This type of erroneous classification is however unlikely in the case study sample given the selected projects had relatively large team sizes and that relatively few women participate in open source projects on GitHub.

\section{Conclusion}

In this paper, we reported the results of a quantitative study of gender differences in membership change in open source projects hosted on GitHub, a social coding platform. 
Replicating and extending prior work on gender differences in tenure, we observed that women had shorter tenure when compared to men, and that unknown gender contributors similarly had shorter tenure in the platform. 
Our application of Nagappan et al.'s \cite{nagappan_diversity_2013} sample coverage scoring approach provides evidence that the open source projects maintained by mixed-gender teams are not quantitatively distinct from those that are maintained by teams who do not have a mixed-gender composition. 
The results presented here also suggest the need to further explore the relationship between gender and status and expertise in open source projects, as we found that these factors, in addition to activity levels, predicted turnover. 
Future work in this area may reveal important insights about the experiences of women as they relate to power dynamics and membership change in open source projects. 
Lastly, we observe that, although the overall project team is relatively stable, there are gender differences in the movement in and out of open source projects.

\bibliographystyle{IEEEtran}
\bibliography{IEEEabrv,bibbybib}

\vspace{12pt}

\end{document}